\newcounter{myctr}

\documentclass{ws-acs}

\usepackage{url}

\begin{document}

\markboth{Cappellini and Ferraris}{Waiting Times in Simulated Stock Markets}

%
\catchline{}{}{}{}{}
%

\title{Waiting Times in Simulated Stock Markets}

\author{\footnotesize ALESSANDRO N. CAPPELLINI}

\address{ISI Foundation, Corso Settimio Severo 65\\
Torino, Italy\\
cappellini@econ.unito.it}

\author{GIANLUIGI FERRARIS}

\address{University of Torino, Corso Unione Sovietica 218\\
Torino, Italy\\
ferraris@econ.unito.it}

\maketitle

\begin{history}
\received{(received date)}
\revised{(revised date)}
\end{history}

\begin{abstract}
Exploiting a precise reproduction of a stock exchange, the robustness of the Continuous Double Auction (CDA) mechanism, evaluated by means of the waiting time distributions, has been proved versus 36 different set ups made by varying both the operators' behaviour and the market micro structure. The obtained results demonstrate that the CDA remains able to clear strongly different order flows, though the Milan stock exchange seemed to be a little more efficient than the NYSE under the allocative point of view, witnessing the intrinsic complexity of the stock market. The simulation has been built as an Agent Based Model in order to  obtain a plausible order flow. The decisions of single agents and their interaction through the market book  are realistic and reproduce some empirical analysis results.
The mentioned results have been obtained either by the analysis of the complete pending time series and the same computation of the asks and bids series alone.

\end{abstract}


\section{Introduction}
\label{sec:Introduction}

Stock market is pretended to be a complex system where small variation in the environment could dramatically affect the outcomes.  This research focuses on our own opinion that Continuous Double Auction (CDA) seems to show the same behavior in waiting time distribution even under different setups and different order flows. 

To investigate this idea we used an agent based model able to reproduce both the Milan stock exchange and the NYSE orders rankings. Into the model operate a wide bunch of fully similar agents that simply toss random decisions about: i) buying or selling, ii) the share quantity going to be bought or sold and iii) the requested price. To create different orders flows, various decision routines are used.

During the research several different market scenarios have been experimented through the interaction of  one thousand agents for one hundred days each time. Each experiment has been repeated using different pseudo random generation seeds to check the obtained figures. The observation of such data confirms that  the waiting time distribution seem to be fully independent by both different setups and order flows,  so it appears to be only related to the continuous double action mechanism.

\section{Book setups and Orders Flows Modeling}
\label{sec:BooksetupsandOrdersFlowsModelling}

Many regulated markets implement trading via the continuous double auction (CDA),
through which buyers and sellers send their orders at any times. These orders, including price
and quantity (volume), are collected into a book. The limit order book is designed for the continuous double auctions mechanism.  
In a CDA,  orders may be submitted in any moment during the trading period. At any time if there are open bids
and asks that are compatible in terms of price and quantity
of goods, a trade is executed immediately. 
Then some information about the trade are broadcast to all participants (price, volume...) depending on local rules.

Depending on market regulations, traders may
submit various kinds of orders. The most common are limit orders: orders to buy a stated amount of a security at or below a given price, or to sell it at or above a given price (the limit price); There are also market orders: orders to buy or sell a stated quantity of a security at the most advantageous price obtainable after the order is represented in the Trading Crowd.

\subsection{The Limit Order Book}
\label{sec:TheLimitBook}

The limit order book of a specific stock, at a specific instant in time $t$, can be described as follows:

\begin{equation}
\beta_n \leq \ldots \beta_3 \leq  \beta_2 \leq \beta_1 < \alpha_1 \leq \alpha_2 \leq \alpha_3 \ldots \leq \alpha_m
\label{eq:book}       
\end{equation}
where the $\beta_i$ represent buy limit orders (bids) and the $\alpha_i$ sell limit orders (offers or asks); those limit orders are characterized by a limit price, a volume and a time of arrival in the book, and they are all waiting in queue to be (potentially) executed. The highest bid $\beta_1$, also called best bid, and the lowest offer $\alpha_1$, or best offer, define the spread $\alpha_1-\beta_1$. Orders queued in the book are usually sorted by price, time of arrival and volume, with variations from market to market -- see \cite{hasbrouck0} for a more detailed description. $\beta_1$ will be executed, only if the book receives a market sell order or a limit sell order with an offering price lower than $\beta_1$ ($\beta_1\ge \alpha_x$). In this case, a trade is generated and the new market price becomes $\beta_1$.

But, can changes in rules/architectures  really affect market performances?

Some researchers \cite{bottazzi1,PellizzariDalForno2005,LiCalziPellizzari2006} compared 
price dynamics of different market protocols (Walrasian market, batch
auction, continuous double auction and dealership) using  agent-based
artificial exchanges. 

We introduce a comparison of rules for  incoming orders flows ranking in the book, according to the Price-Time-Quantity rule and the Price-Quantity-Time one\footnote{
This comparison was firstly overviewed in a previous work \cite{cappellini2006}, founding that:
\begin{quotation}
	In a homogeneous population of traders there is no concrete advantage regarding the two rules, as shown by our results for which Milan is a little quick in order execution.
	However in real world orders are strategically placed for every level of prices, that means to be quicker (PTQ) or bigger (PQT) than the average of traders.
\end{quotation}
}.

The Price-Quantity-Rule (herefore ``Nyse'', PQT or NY) is used in New York Stock Exchange\footnote{\url{http://rules.nyse.com/}}:

\begin{quotation}
{\bf Rule 72. Priority and Precedence of Bids and Offers}

	{\bf I. Bids.} -- Where bids are made at the same price, the priority and precedence shall be determined as follows:
	
	{\bf Priority of first bid}
	
	(a) Except as provided in paragraph (b) below, when a bid is clearly established as the first made at a particular price, the maker shall be entitled to priority and shall have precedence on the next sale at that price, up to the number of shares of stock or principal amount of bonds specified in the bid, irrespective of the number of shares of stock or principal amount of bonds specified in such bid.
	
	[...]
\end{quotation}

The Price-Time-Quantity rule (herefore ``Milan'', PTQ or MI) is applied in various order-driven markets like the Italian Stock Exchange\footnote{\url{http://www.borsaitaliana.it/documenti/regolamenti/}}:

\begin{quotation}
{\bf Article 4.1.4}

	The orders for each instrument shall be automatically ranked on the book according to price
	-- in order of decreasing price if to buy and increasing price if to sell -- and, where the price
	is the same, according to entry time. Modified orders shall lose their time priority if the
	modification implies an increase in the quantity or a change in the price.
	
	[...]
\end{quotation}

\subsection{Orders Flows Modeling}
\label{sec:OrdersFlowsModelling}

In the stock market simulations,   the basic price decision mechanism is usually very simple, the agents:

\begin{enumerate}
	\item  know only the last executed price;
	\item  choose randomly the buy or sell side;
	\item  fix their limit price and quantity.
\end{enumerate}

 Usually the order price at time $t$ ($o_t$) is defined as a variation of the last executed price in the market ($p_{t-1}$).

The orders price can be defined as a Gaussian (G),

\begin{equation}
o_t = N(p_{t-1},0.2)
\label{eq:guassian}
\end{equation}

or as a multiplicative process

\begin{equation}
o_t = p_{t-1} \cdot \xi
\label{eq:multiplicative}
\end{equation}

where $\xi$ can be $ N(1,0.2)$ (MG) or $U[0.5;1.5]$ (MU),

or additive

\begin{equation}
o_t = p_{t-1} + \xi
\label{eq:additive}
\end{equation}

where $\xi$ can be   $N(0,0.2)$ (AG) or $U[-1;1]$ (AU),

or exponential (E)

\begin{equation}
o_t = p_{t-1} \cdot e^{\delta \cdot \xi}
\label{eq:exponential}
\end{equation}

where $\delta$ is the price deviation, 0.02 and $\xi = N(0,1)$.

The order quantity $q_t$ can be modeled according to a Gaussian (G),
 $N(2,50)$, or an uniform distribution (U), $U[1;100]$, or can be simply constant and equal to $1$ (S).

For this work, we explore all the combination of prices and quantity decisions, having 18 scenarios to be applied to the two different book mechanisms. They are labeled according to previous described acronyms, so ``AU G MI'' means we adopted an additive uniform price process, a Gaussian size and the Milan orders ranking.

\section{The simulation framework: SumWEB}
\label{sec:ThesimulationframeworkSUM}

The Agent Based Modeling (ABM) paradigm \cite{Testfatsion} focuses on the central role of the agent, the representation in silicon of a real operator able to perform an own and autonomous behavior. Artificial agents may be endowed with the ability to compute and modify their own strategies to adapt them to the current conditions of the environment. Following the ABM approach, phenomena observed at an aggregate level could be reproduced through the interaction among simple entities performed  by running the simulation. ABM has  become the main instrument to deal with complexity, so it is strongly eligible as a tool to investigate matters related to the evolution and adaptation of autonomous behaviors.

SumWEB (SUM Web Economic Behaviour)\footnote{\url{http://eco83.econ.unito.it/sumweb/} described in \cite{cappellini2003} and \cite{sistemiintelligenti2005}} is basically an extension of Pietro Terna's \cite{terna00,terna00a} simulation SUM (Surprising (Un)realistic Market).

The main feature of the software is the 
realism of  implementation. The large part of the coded mechanisms were directly inspired by MTA (\textit{Mercato Telematico Azionario}, the Italian Stock Exchange), that's an order driven book. So SUM and SumWEB have a
tick per tick realistic formation price mechanism (a continuous double auction CDA), with characteristics as opening and closing auctions, market and limit price orders.

Then it  can manage more than one single stock, and different financial instruments (futures and market indexes). This means that it can be used to understand rules and regulation of real markets dealing with their micro-foundations, or to act as policy maker developing and new rules tester.

The Sum model can host several type of artificial agents:
\begin{itemize}
	\item Random. They are the simplest agents, that choose randomly their actions. They represent a crowd of little investors;
	\item Financial Technique. They use schemata and rules such as traders. Eg. They can arbitrage a future contract against the underlying assets or use stop-loss/take-profit strategy;
	\item Simple Cognitive. They perform simple behavior observing the market operations. They can imitate the last movement or ask for suggestion to other agents;
	\item Social. Agents  increase their knowledge by ``informal'' chatting into the group and, after pondering, raise beliefs to be used both acting in the market and spreading to others. 
	\item Minded. They have a  cognitive-like and adaptable structure. They are principally genetic algorithms, neural networks and classifier systems. They evolve themselves during every run;
	\item Avatar, the interface for humans. They ``capture'' humans' orders bringing them into the book.
\end{itemize}

As technological overview of the software we can observe that it is a SWARM\footnote{\url{http://www.swarm.org}} \cite{swarm} based Objective-C simulation.

\section{Results}
\label{sec:Results}

\begin{figure}[htbp]
 \centerline{\includegraphics[width=0.8\textwidth]{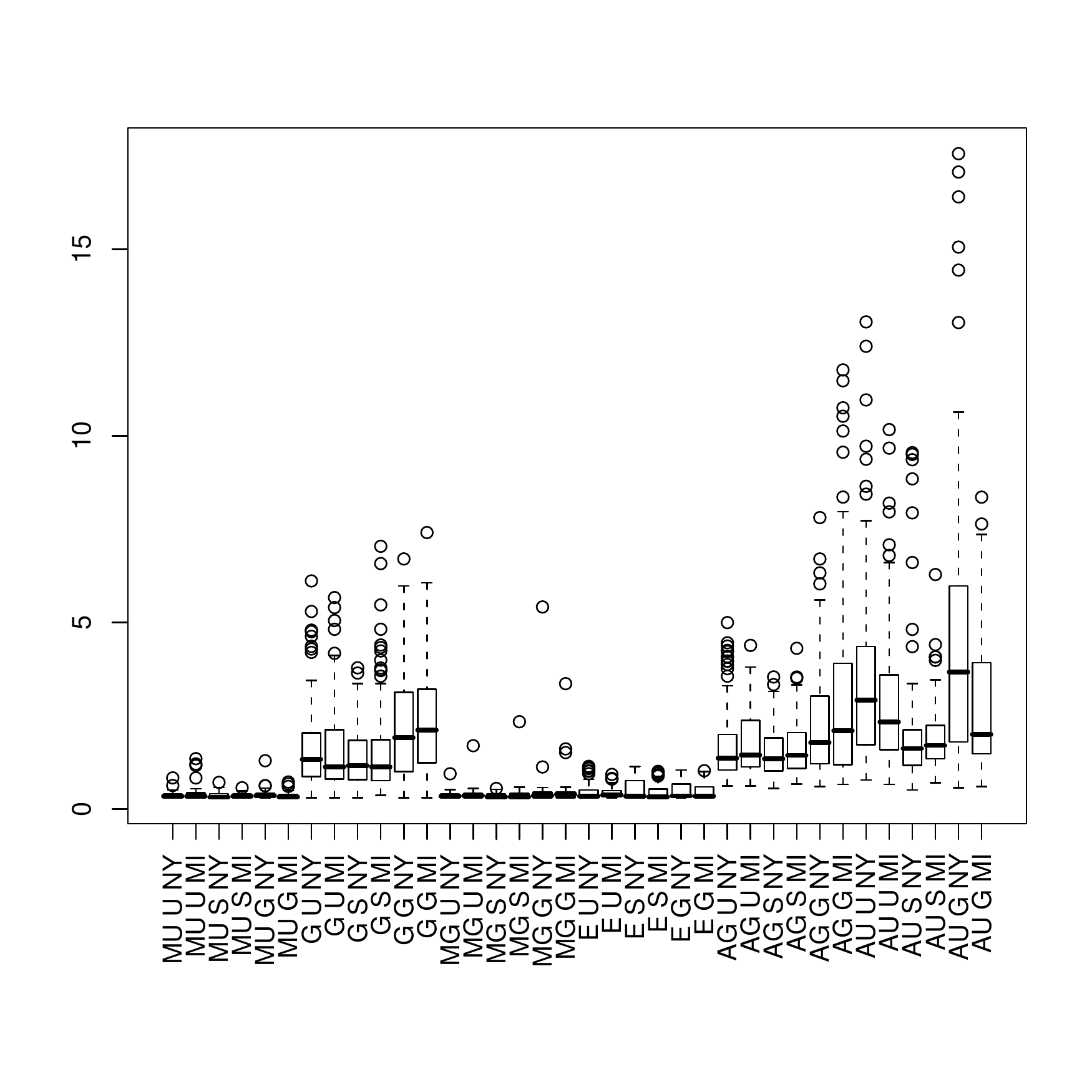}}
 \caption{Closing Prices box plots for several realizations}
 \label{fig:closingPrices}
 \end{figure}

We ran several simulations for each scenario, composed by different decisions mechanisms in term of quantity or price, and by different market rules. The price realizations are quite heterogeneous, as shown by box plot (figure \ref{fig:closingPrices}) of closing prices for different scenarios.

\begin{figure}[htbp]
 \centerline{\includegraphics[width=0.8\textwidth]{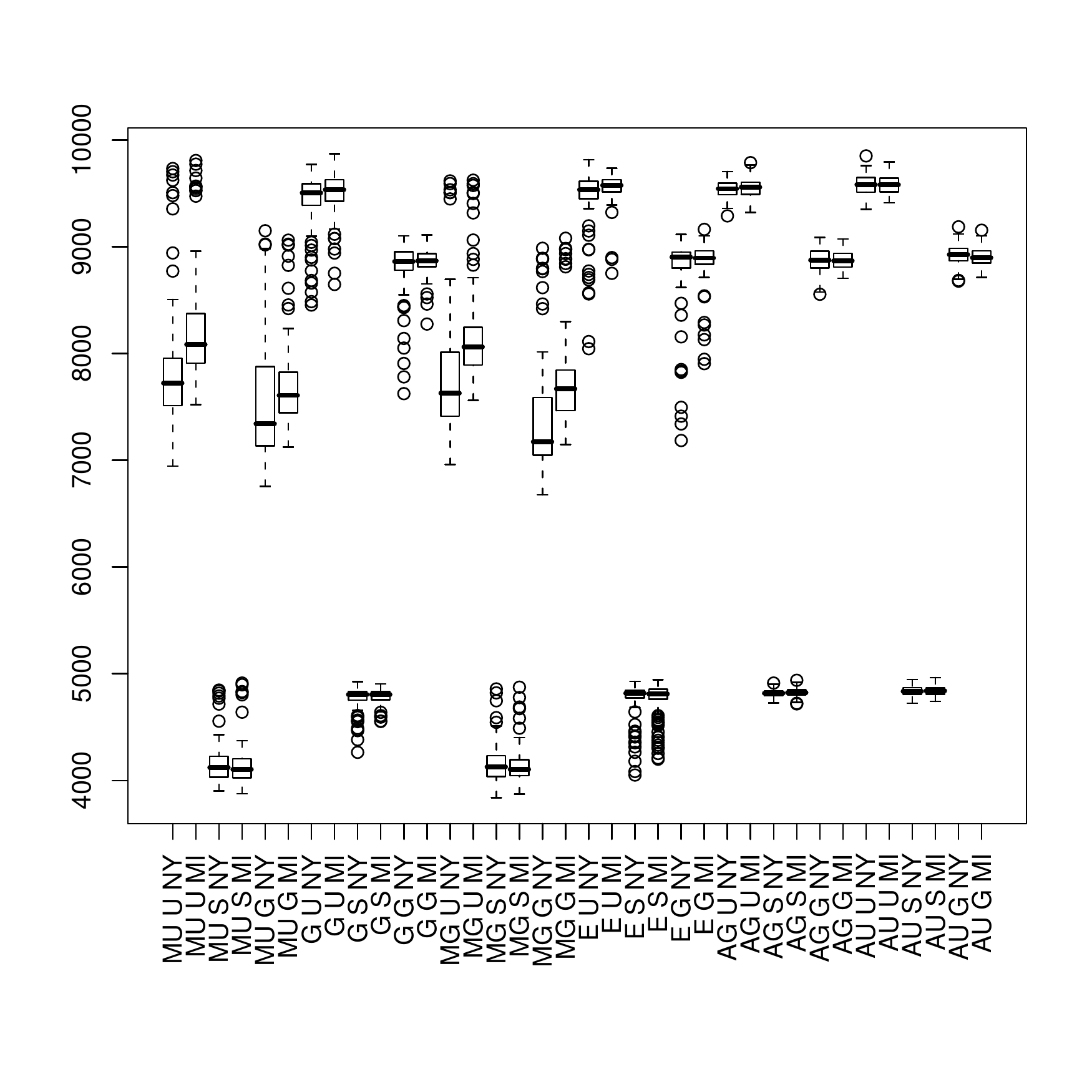}}
 \caption{Trades numbers box plots for several realizations}
 \label{fig:tradesNumbers}
 \end{figure}

All the realizations show the same shapes in waiting (pending) time distribution (see \cite{scalas2006,scalas2002,scalas2004}) under different market conditions. 

We define tick time as the arrival time in book of every order (incremented by one at every event), so the pending time\footnote{The waiting times could be defined as follows:

\begin{itemize}
	\item intertrade time, among executed orders;
	\item arrival time, among orders submissions;
	\item pending time, permanence in book until their execution.
\end{itemize}

The tick times is an internal measure of the simulation, regarding events, such as every agent actions or book events (executions, submissions, etc.). In this work we use the orders cardinality, or the progressive number of each order as they were received by the book.}
is the amount of time an order waits in the book until its execution.

Here we explore the result of 1000 agents operating for 100 days of 12 ``turns/hours'', into a stock market that applies the ranking rules of the NYSE (PQT). The agents use a multiplicative price driven by a uniform distribution of prices (eq. \ref{eq:multiplicative}) and a uniform distribution for quantity.

\begin{figure}[htbp]
 \centerline{\includegraphics[width=0.95\textwidth]{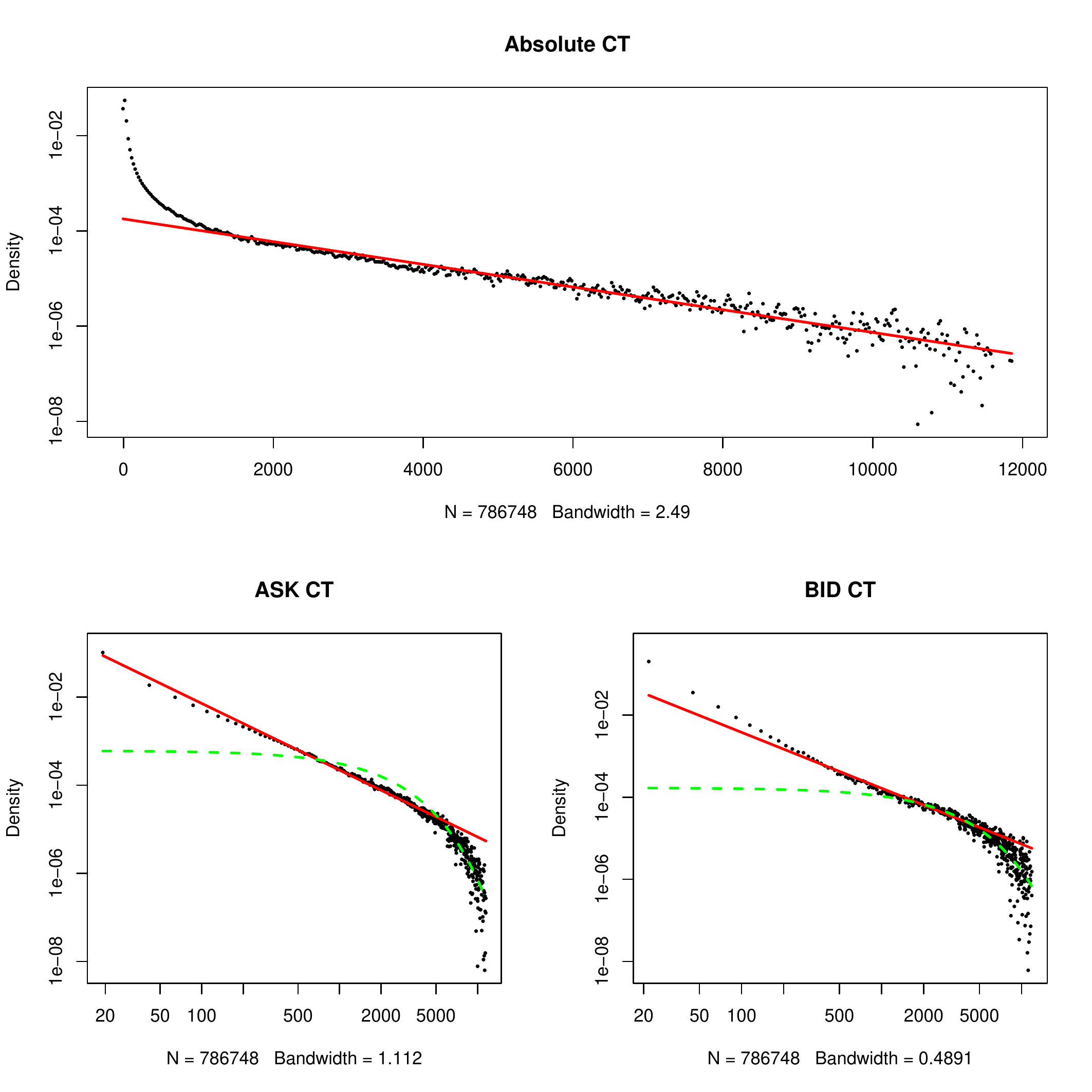}}
 \caption{Pending Time probability distribution for PQT market having orders with multiplicative-uniform prices and uniform quantities}
 \label{fig:UniformUniformNyse-fitting}
 \end{figure}

We first analyze the complete series of pending times (``absolute'') during continuous trading. The main result is a good exponential fitting, drawn in red solid line in figure \ref{fig:UniformUniformNyse-fitting}. 
The $\alpha$\footnote{$y = A e^{\alpha x}$ The $A$ parameter is -3.748, R-squared 0.9429. The p-values are 99\% siginificant for both parameters.} exponent
is -2.383e-04.

Through a linear regression on the log-log plot, we estimate a power law fitting for Bids and Asks series according to the standard equation: $y = A x^\alpha$ (red solid line).
The $\alpha$\footnote{The $A$ parameter is 0.3026 for Bids and 0.88721 for Asks, R-squared are 0.9723 and 0.9859, respectively. The p-values are 99\% siginificant.} exponents are
 -1.36148 for Bids and -1.51508 for Asks. It was quite interesting fit the tails of Asks and Bids distribution with an exponential (green dashed line), which $alpha$ are -2.023e-04 (Bids) and -2.921e-04 (Asks)\footnote{R-squared 0.6347 and 0.7315, intercept -3.774 and -3.218}.
These exponents were confirmed by empirical analysis in \cite{challet2001}, who found a typical value in $-1.5 \pm 0.1$.

In figure \ref{fig:tradesNumbers} the trades number distributions across the different scenarios are drawn. Obviously at the bottom are those scenarios used a single share per order (S). The others show different patterns, even if the standard descriptive statistics figures (average, median...) underline that the PTQ ranking rule is a little bit more allocative efficient than PQT rule adopted in New York.

\section{Criticism}
\label{sec:Criticism}

We found very similar shapes for all scenarios, but we would like to underline two limitations: the poorness of traders population and the  market rules not yet fully explored.

To keep the focus on market mechanisms, we decide to simplify agents behaviour using  random agent only, that don't perform any market analysis or any valuation on their own portfolio. So our agents act at any turn and cannot wait for strategically issuing orders, they are not sensible to any kind of news and events, and they do not relax neither sleep or simply wait as in \cite{MuchnikSolomon2006}.

The implementation of CDA mechanism is robust and flexible. In the presented realizations it still miss some details.

Orders cancellation, market suspensions and circuit breakers  can increase the realism but we still prefer to follow the simplest market implementation.

\begin{figure}[htbp]
 \centerline{\includegraphics[width=0.8\textwidth]{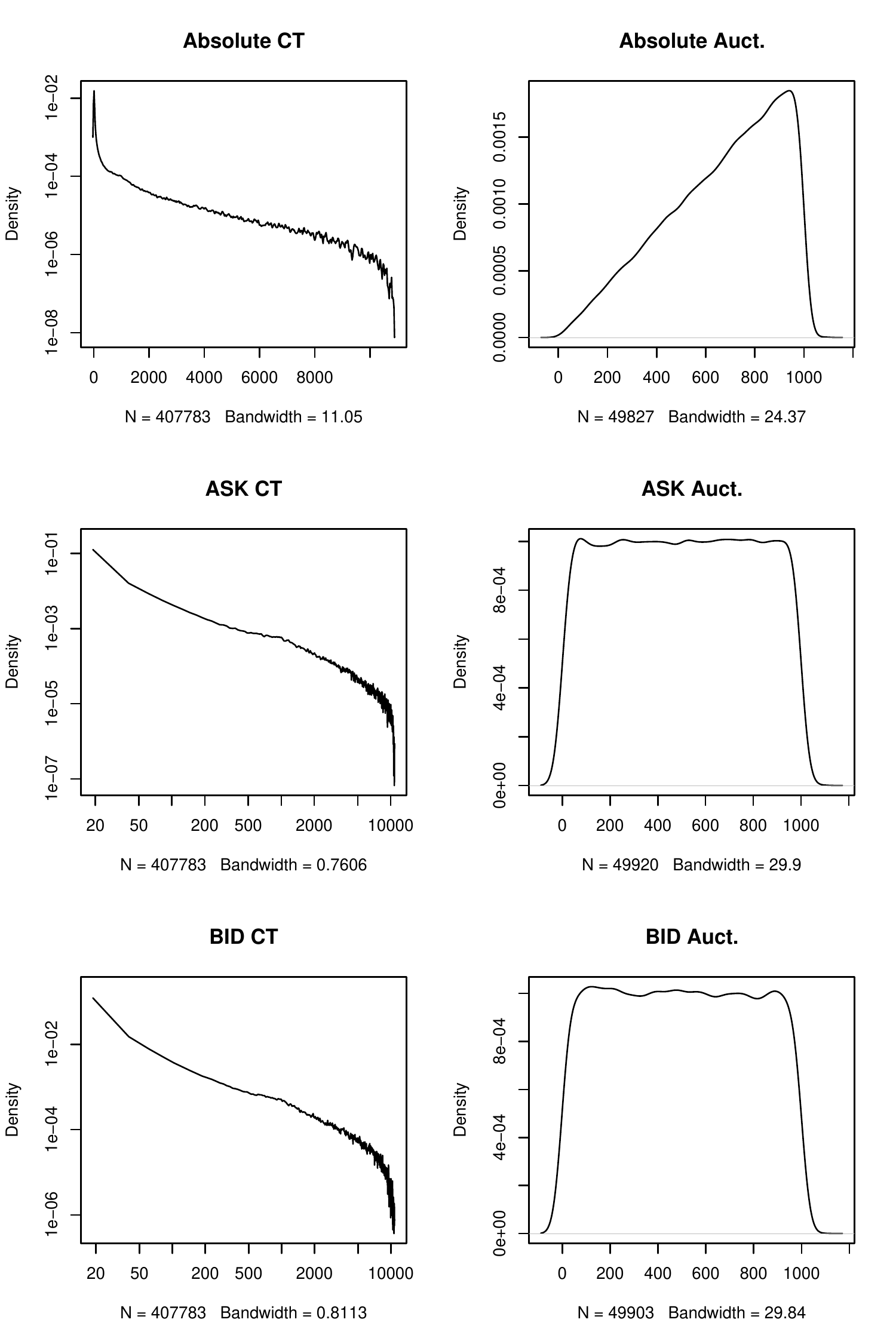}}
 \caption{Continuous Time and Auction pending times probability for PTQ market having orders with exponential prices and single quantities}
 \label{fig:auction}
 \end{figure}

As example the figure \ref{fig:auction} shows a Milan like market (PTQ), populated by 1000 agents adopting exponential prices (eq. \ref{eq:exponential}) with single quantities, for 100 days. In this simulation we add the opening and the closing auctions\footnote{The auction is modeled according to Milan Stock Exchange rules.} at the beginning and at the end of all days.

 During auction time the orders are collected and enqueued into the book until the auction starts; then the auction price is set up and the orders are matched as far as possible according to their rank. So, we consider the ask (or bid) pending time as the time they wait until the execution (at current time $t$) into the book ($\tau_{ask} = t - t_{ask}$) while ``absolute'' time is the longest time the orders waited into the book ($\tau_{abs} = t - \min(t_{bid} ; t_{ask})$).
 
 Those results confirm  the general shapes for continuous time executed order probability distribution (exponential for absolute, and power law for ask and bid) and show very particular behaviours: triangular probability for absolute prices, and flat for asks and bids.
The latest effects are due to a constant auction time at $t=1000$  and a global uniform probability for orders to be issued in a time $[0;1000]$.

\section{Conclusions}
\label{sec:Conclusions}

The  results witnessed the existence of some  regularities that may be interpreted as the emergence of a property related to the continuous double auction mechanism. This property has shown itself to be independent from the different rules used to enqueue incoming orders in Milan stock exchange (so called PTQ) and NYSE (PQT), as well as from diverse random distributions adopted to determine the behaviors of the agents  in setting up both prices and quantities.

A possible reason of this emergence could be related to the capability of the indirect interaction among the agents to clean up the waiting queues, mainly due to the prices trend. The agents are always changing their requested prices by slightly biasing current ones, such a mechanism generated  a prices dynamic that ensures the large majority of the order founded a counterpart in a reasonable  time span.

The results could be classified as emergent because, in a priori reasoning there are not strong reason for expecting that independent random walking agents will generate a decisions time sequence that systematically would nearly clean up the market.

\vspace*{-3pt}   



\vspace*{-5pt}   


\begin{thebibliography}{10}

\bibitem{bottazzi1}
G.~Bottazzi, G.~Dosi, and I.~Rebesco.
\newblock Institutional architectures and behavioral ecologies in the dynamics
  of financial markets.
\newblock {\em Journal of Mathematical Economics}, 41(1-2):197--228, 2005.

\bibitem{cappellini2003}
Alessandro Cappellini.
\newblock Esperimenti su mercati finanziari con agenti naturali ed artificiali.
\newblock Master's thesis, Facoltà di Economia, Torino, July 2003.

\bibitem{sistemiintelligenti2005}
Alessandro Cappellini.
\newblock Avatar e simulazioni.
\newblock {\em Sistemi intelligenti}, (1):45--58, Aprile 2005.

\bibitem{cappellini2006}
Alessandro Cappellini.
\newblock {\em Heterogeneous approaches to financial markets}.
\newblock PhD thesis, University of Torino, Italy, November 2006.

\bibitem{challet2001}
D.~Challet and R.~Stinchcombe.
\newblock Analyzing and modeling 1+1d markets.
\newblock {\em Physica A: Statistical Mechanics and its Applications},
  300:285--299, November 2001.

\bibitem{hasbrouck0}
J.~Hasbrouck, G.~Sofianos, D.~Sosebee, and F.~Director.
\newblock New york stock exchange systems and trading procedures, 1993.

\bibitem{swarm}
C.~Langton, N.~Minar, and R.~Burkhart.
\newblock Swarm web site, 1995.
\newblock \url{http://wiki.swarm.org}.

\bibitem{LiCalziPellizzari2006}
Marco LiCalzi and Paolo Pellizzari.
\newblock The allocative effectiveness of market protocols under intelligent
  trading.
\newblock Working Papers 134, Department of Applied Mathematics, University of
  Venice, May 2006.

\bibitem{MuchnikSolomon2006}
L.~Muchnik and S.~Solomon.
\newblock Markov nets and the natlab platform; application to continuous double
  auction.
\newblock In M.~Salzano and D.~Colander, editors, {\em Complexity Hints for
  Economic Policy}. Springer-Verlag, Berlin, 2007.

\bibitem{PellizzariDalForno2005}
Pellizzari, P., Forno, A.D.:
\newblock A comparison of different trading protocols in an agent-based market.
\newblock Journal of Economic Interaction and Coordination \textbf{2}(1) (2007)
   27--43
   
\bibitem{scalas2002}
Marco Raberto, Enrico Scalas, and Francesco Mainardi.
\newblock Waiting-times and returns in high-frequency financial data: an
  empirical study.
\newblock {\em Physica A: Statistical Mechanics and its Applications},
  314:749--755, November 2002.

\bibitem{scalas2004}
Enrico Scalas, Rudolf Gorenflo, Hugh Luckock, Francesco Mainardi, Maurizio
  Mantelli, and Marco Raberto.
\newblock Anomalous waiting times in high-frequency financial data.
\newblock {\em Quantitative Finance}, 4(6):695--702, December 2004.

\bibitem{scalas2006}
Enrico Scalas, Taisei Kaizoji, Michael Kirchler, Jürgen Huber, and Alessandra
  Tedeschi.
\newblock Waiting times between orders and trades in double-auction markets.
\newblock {\em Physica A: Statistical Mechanics and its Applications},
  366:463--471, July 2006.

\bibitem{terna00a}
Pietro Terna.
\newblock Cognitive agents behaving in a simple stock market structure.
\newblock In F.~Luna and A.~Perrone, editors, {\em Agent-Based Methods in
  Economics and Finance: Simulations in Swarm}. Kluwer Academic, Dordrecht and
  London, 2000.

\bibitem{terna00}
Pietro Terna.
\newblock Sum: a surprising (un)realistic market: Building a simple stock
  market structure with swarm.
\newblock 2000.
\newblock CEF 2000, Barcelona, June 5-8.

\bibitem{Testfatsion}
L.~Testfatsion.
\newblock Agent-based computational economics.
\newblock Technical Report~1, ISU Economics working Paper, 2003.

\end{thebibliography}


\end{document}